\documentclass{biometrika}

\usepackage{amsfonts,amsmath}
\usepackage{natbib,graphicx}
\usepackage{times}

\usepackage{bm}

\newcommand{\E}{E}

\newcommand\Evid{Z}
\newcommand{\eps}{\varepsilon}

\begin{document}

\title{\bf Properties of nested sampling
} 

\author{Nicolas Chopin}\affil{CREST--ENSAE, Timbre J120, 
3, Avenue Pierre Larousse, 92245 Malakoff cedex, France\email{nicolas.chopin@ensae.fr}}
\author{and Christian P.~Robert}\affil{CEREMADE, Universit\'e Paris Dauphine, F-75775 Paris cedex 16, France\email{xian@ceremade.dauphine.fr}}

\maketitle

\begin{abstract}
Nested sampling is a simulation method for approximating marginal
likelihoods proposed by \cite{Skilling:2007a}. We establish
that nested sampling has an approximation error that vanishes at the standard Monte Carlo rate 
and that this error is asymptotically Gaussian. We show that the asymptotic variance of the nested sampling approximation
typically grows linearly with the dimension of the parameter. We 
discuss the applicability and efficiency of nested sampling in realistic
problems,  and we compare it with two current methods for computing marginal likelihood. 
We propose an extension that avoids resorting to Markov chain Monte Carlo to obtain the simulated points.
\end{abstract}


\section{Introduction}

Nested sampling was introduced by \cite{Skilling:2007a} as a numerical 
approximation method for integrals of the kind
\[
Z = \int L(y\mid \theta) \pi(\theta)\,\text{d}\,\theta\,,
\]
when $\pi$ is the prior distribution and $L(y\mid \theta)$ is the likelihood.
Those integrals are called evidence in the above papers. They naturally occur 
as marginals in Bayesian testing and model choice (\citealp{jeffreys:1939}; 
\citealp{Robert:2001}, Chapters 5 and 7). Nested sampling has been well received in
astronomy and has been applied successfully to several cosmological problems,
see, for instance, \cite{mukherjee2006nsa}, \cite{shaw2007ebi}, and \cite{Vegetti}, among others.
In addition,  \cite{Murray:nested:Potts} develop a nested sampling algorithm for
computing the normalising constant of Potts models. 

The purpose of this paper is to investigate the formal properties of nested sampling. 
A first effort in that direction is \cite{evans:2007}, which shows that 
nested sampling estimates converge in probability, but calls for further work on
 the rate of convergence and the limiting distribution. 
 
Our main result is a central limit theorem for nested sampling estimates, which says that 
the approximation error is dominated by a  $\text{O}(N^{-1/2})$ stochastic term, which 
has a limiting Gaussian distribution, and where $N$ is a tuning parameter proportional 
to the computational effort. 
We also investigate the impact of the dimension $d$ of the problem on the performances of 
the algorithm. In a simple example, we show  that the asymptotic variance of nested sampling 
estimates grows linearly with $d$; this means that the 
computational cost is $\text{O}(d^3/\eta^2)$, where $\eta$ is the selected error bound. 

One important aspect of nested sampling is that it resorts to simulating  
points $\theta_i$ from the prior $\pi$, constrained to $\theta_i$ having a larger
likelihood value than some threshold $l$. In many cases, the simulated points must be
generated by Markov chain Monte Carlo sampling. We propose an extension of nested sampling, based
on importance sampling, that introduces enough flexibility so
as to perform the constrained simulation without resorting to Markov chain Monte Carlo.

Finally, we examine two alternatives to nested sampling for computing evidence, both based 
on the output of Markov chain Monte Carlo algorithms. We do not aim at an exhaustive comparison with all
existing methods, see, for instance, \cite{chen:shao:ibrahim:2000}, for a broader review,
and restrict our attention to methods that share the property with nested sampling
that the same algorithm provides approximations of both the posterior 
distribution and the marginal likelihood, at no extra cost. 
We provide numerical comparisons between those methods, since some of the aforementioned papers and 
Murray's PhD thesis (2007, University College London), also include numerical comparisons of nested sampling with 
other methods for several models.

\section{Nested sampling: A description}
\subsection{Principle}
We briefly describe the nested sampling algorithm, as introduced by  \cite{Skilling:2007a}. We use $L(\theta)$ as
a short-hand for the likelihood $L(y\mid \theta)$, omitting the dependence on $y$. 

Nested sampling is based on the following  identity: 
\begin{equation} \label{eq:onedimint}
 Z = \int_0^1 \varphi(x)\,\text{d}x\,,
\end{equation}
where $ \varphi$ is the inverse of the survival function of the random variable $L(\theta)$,
\[\varphi^{-1}:l\rightarrow \mathrm{pr}\{L(\theta)>l\}\,,\]
 assuming $\theta\sim\pi$ and $\varphi^{-1}$ is a 
decreasing function, which is the case when $L$ is a continuous function
and $\pi$ has a connected support. The representation
$\Evid=E^\pi\{L(\theta)\}$ holds with no restriction on either $L$ or $\pi$.
Formally, this one-dimensional integral could be approximated by standard quadrature methods, 
\begin{equation} \label{eq:RiemannSum1} 
\widehat{Z} = \sum_{i=1}^{j}(x_{i-1}-x_{i})\varphi_i\,,
\end{equation}
where $\varphi_i=\varphi(x_i)$, and $0<x_j<\cdots<x_1<x_0=1$ is an arbitrary grid over $[0,1]$.
Function $\varphi$ is intractable in most cases however, so the $\varphi_i$'s are approximated by 
an iterative random mechanism:
\begin{itemize}
\item[--] Iteration 1: draw independently $N$ points $\theta_{1,i}$ from the prior $\pi$, 
determine
$
\theta_1 = \arg \min_{1\le i\le N} L(\theta_{1,i}),
$
and set $\varphi_1=L(\theta_1)$. 
\item[--] Iteration 2: 
obtain the $N$ current values $\theta_{2,i}$, by reproducing the 
$\theta_{1,i}$'s, except for $\theta_1$ that is
replaced by a draw from the prior distribution $\pi$ conditional
upon $L(\theta)\ge \varphi_1$; then select $\theta_2$ as 
$
\theta_2 = \arg \min_{1\le i\le N} L(\theta_{2,i}),
$
and set $\varphi_2=L(\theta_2)$. 
\item[--] Iterate the above step until a given stopping rule is satisfied, for instance 
when observing very small changes in the approximation $\widehat{Z}$ or when reaching 
the maximal value of $L(\theta)$ when it is known.
\end{itemize}

In the above, the values $x_i^\star=\varphi^{-1}(\varphi_i)$ that should be used 
in the quadrature approximation (\ref{eq:RiemannSum1}) are unknown, but they have 
the following property: 
$
t_i=\varphi^{-1}(\varphi_{i+1})/\varphi^{-1}(\varphi_i) = x_{i+1}^\star/x_i^\star
$ 
are independent $\textrm{beta}(N,1)$ variates.
\cite{Skilling:2007a} proposes two approaches: first, a deterministic scheme, where $x_i$
is substituted with $\exp(-i/N)$ in (\ref{eq:RiemannSum1}), so that $\log x_i$ is the expectation of 
$\log \varphi^{-1}(\varphi_i)$; second, a random scheme, where $K$ parallel streams of random
numbers $x_{i,k}$, $k=1,\ldots,K$, are generated from the same generating process
as the $x_i^\star$, $x_{i+1,k}=x_{i,k} t_{i,k}$, where $t_{i,k}\sim \textrm{beta}(N,1)$. 
In the latter case, a natural estimator is:
$$ 
\log \widetilde Z = \frac 1 K \sum_{k=1}^K \log  \widetilde Z_k,\quad  
\widetilde Z_k= \sum_{i=1}^j (x_{i-1,k}-x_{i,k})\varphi_i\,.
$$

For the sake of brevity, we focus on the deterministic scheme in this paper, and study the 
estimator (\ref{eq:RiemannSum1}) and $x_i=\exp(-i/N)$. Furthermore, for $K=1$, the random scheme produces
more noisy estimates than the deterministic scheme, but, for large values of $K$, it may be the opposite,
see for instance Fig.~3 in \cite{Murray:nested:Potts}.

\subsection{Variations and posterior simulation} \label{variat:sec}


\cite{Skilling:2007a} indicates that nested sampling 
provides simulations from the posterior distribution at no extra cost:
``the existing sequence
of points $\theta_1,\theta_2,\theta_3,\ldots$ already gives a set of posterior representatives,
provided the $i$'th is assigned the appropriate importance weight $\omega_i L_i$", 
where the weight $\omega_i$ is equal to the difference $(x_{i-1}-x_i)$ and $L_i$ is equal to $\varphi_i$.
This can be justified as follows. Consider the computation of the posterior expectation
of a given function $f$ 
\[
\mu(f) = {\int \pi(\theta)L(\theta)f(\theta)\,\text{d}\theta}
\bigg/{\int \pi(\theta)L(\theta)\,\text{d}\theta}\,.
\]
One can then use a single run of nested sampling to obtain estimates of 
both the numerator and the denominator, the latter being the evidence $\Evid$,
estimated by (\ref{eq:RiemannSum1}). The estimator
\begin{equation}\label{estnum:eq}
\sum_{i=1}^{j}(x_{i-1}-x_{i})\varphi_i f(\theta_i)  
\end{equation}
of the numerator is a noisy version of 
\[
\sum_{i=1}^{j}(x_{i-1}-x_{i})\varphi_i \widetilde{f}(\varphi_i)\,, 
\]
where $\widetilde f(l)=\E^\pi\{f(\theta)\mid L(\theta)=l\}$,
the prior expectation of $f(\theta)$ conditional on $L(\theta)=l$. This
Riemann sum is, following the principle of nested sampling, an estimator of
the evidence.
%
\begin{lemma} \label{post:Lemma}
Let $\widetilde f(l)=\E^\pi\{f(\theta)\mid L(\theta)=l\}$~for $l>0$, then, if
$\widetilde{f}$ is absolutely continuous,
\begin{equation}\label{eq:postequ}
\int_0^1 \varphi(x) \widetilde{f}\{\varphi(x)\}\,\mathrm{d}x = 
\int \pi(\theta)L(\theta)f(\theta)\,\mathrm{d}\theta. 
\end{equation}
\end{lemma}

A proof is provided in Appendix 1. Clearly, the estimate of $\mu(f)$
obtained by dividing (\ref{estnum:eq}) by (\ref{eq:RiemannSum1}) is the estimate 
obtained by computing the weighted average mentioned above. We do not discuss further this aspect
of nested sampling, but our convergence results can be extended
to such estimates. 

\section{A central limit theorem for nested sampling}\label{CLT}

We decompose the approximation error of nested sampling as follows:
\begin{align*}
\sum_{i=1}^{j}(x_{i-1}&-x_{i}) \varphi_i-\int_{0}^{1}\varphi(x)\, \text{d}x  
 = -\int_{0}^{\eps}\varphi(x)\, \text{d}x\\
& +\left\{\sum_{i=1}^{j}(x_{i-1}-x_{i})\varphi(x_{i}) 
	-\int_{\eps}^{1}\varphi(x)\, \text{d}x\right\}
  +\sum_{i=1}^{j}(x_{i-1}-x_{i})\left\{\varphi_i -\varphi(x_{i})\right\}.
\end{align*}

The first term is a truncation error, resulting from the feature that
the algorithm is run for a finite time. For simplicity's sake, we assume that 
the algorithm is stopped at  iteration $j=\lceil (-\log\eps)N\rceil $, 
where $\lceil x \rceil $ stands for the smallest integer $k$ such that $x\leq k$, 
so that
$x_{j}=\exp(-j/N)\leq\eps<x_{j-1}$.
More practical stopping rules are  discussed in \S \ref{sec:simus}.
Assuming $\varphi$, or equivalently $L$, bounded from above,
the error $\int_{0}^{\eps}\varphi(x)\, \text{d}x$
is exponentially small with respect to the computational effort.

The second term is a numerical integration error, which, provided
$\varphi'$ is bounded over $[\eps,1]$, is of order $\text{O}(N^{-1})$, since
$x_{i-1}-x_{i}=\text{O}(N^{-1})$.

The third term is stochastic and is denoted
\[
\eta_N=\sum_{i=1}^{j}
	(x_{i-1} -x_{i})\left\{\varphi(x_{i}^{\star})-\varphi(x_{i})\right\}\,,
\]
where the $x_{i}^{\star}$'s are such that $\varphi_i=L(\theta_i)=\varphi(x_{i}^{\star})$, 
therefore $x_{i}^{\star}=\varphi^{-1}(\varphi_i)$.

The following theorem characterises the asymptotic behaviour of $\eta_N$.
\begin{theorem}
Provided that $\varphi$ is twice continuously-differentiable over $[\eps,1]$, and that its two first
derivatives are bounded over $[\eps,1]$, then $N^{1/2}\eta_N$ converges in distribution to a Gaussian distribution
with mean zero and variance
\[
V = -\int_{s,t\in[\eps,1]}s\varphi'(s)t\varphi'(t)\log (s\vee t)\, \mathrm{d}s\,\mathrm{d}t.
\]
\end{theorem}
The stochastic error is of order $\text{O}_{P}(N^{-1/2})$ and it dominates both other error terms. 
The proof of this theorem relies on the functional central limit theorem 
and is detailed in Appendix 2.
A straightforward application of the delta-method shows that the log-scale error, $\log \widehat{Z}  -\log {Z}$, 
has the same asymptotic behaviour, but with asymptotic variance $V/Z^2$. 

\section{Properties of the nested sampling algorithm}
\subsection{Simulating from a constrained prior}\label{Hoax4}
The main difficulty  of nested sampling is to simulate $\theta$  
from the prior distribution $\pi$ subject to the constraint $L(\theta)>L(\theta_i)$;
exact simulation from this distribution is an intractable problem in many realistic set-ups.
It is at least of the same complexity as a one-dimensional slice sampler,
which produces an uniformly ergodic Markov chain when the likelihood $L$ is bounded but may be
slow to converge in other settings \citep{roberts:rosenthal:1999}. 

\cite{Skilling:2007a} proposes to sample 
values of $\theta$ by iterating $M$ Markov chain Monte Carlo steps, using the truncated prior as the 
invariant distribution, and a point chosen at random among the $N-1$ survivors as the starting point.  
Since the starting value is already
distributed from the invariant distribution, a finite number $M$ of iterations produces an
outcome that is marginally distributed from the correct distribution. This however introduces
correlations between simulated points. We stress that our central limit theorem 
applies no longer when simulated points are not independent, and that the consistency 
of nested sampling estimates based on Markov chain Monte Carlo is an open problem. A reason
why such a theoretical result seems difficult to establish is that each iteration involves both
a different Markov chain Monte Carlo kernel and a different invariant distribution. 

There are settings when implementing a Markov chain Monte Carlo move that leaves the 
truncated prior invariant is not straightforward. In those cases, one may instead
implement an Markov chain Monte Carlo move, for instance a random walk Metropolis--Hastings move, 
with respect to the unconstrained prior, and subsample only values that satisfy the
constraint $L(\theta)>L(\theta_i)$, but this scheme gets increasingly inefficient as 
the constraint moves closer to the highest values of $L$.
More advanced sampling schemes can be devised that
overcome this difficulty, such as the use of a diminishing variance factor in
the random walk. 

In \S \ref{sec:nis}, we propose an extension of nested sampling based on
importance sampling. In some settings, this may 
facilitate the design of efficient Markov chain Monte Carlo steps, 
or even allow for sampling independently
the $\theta_i$'s. 

\subsection{Impact of dimensionality}\label{sec:toy}

We show in  this section that the theoretical performance of nested sampling typically depends 
on the dimension $d$ of the problem as follows: the required number 
of iterations
and the asymptotic variance both grow linearly with $d$.
Thus, if a single iteration costs $\text{O}(d)$, the computational cost of 
nested sampling is  $\text{O}(d^{3}/\eta^2)$, where $\eta$ denotes a given error level; 
Murray's PhD thesis also states this result, 
using a more heuristic argument. This result applies to
the exact nested algorithm only. In principle, resorting 
to Markov chain Monte Carlo might entail some additional curse of dimensionality, 
but this point seems difficult to study formally, and will only be briefly investigated
in our simulation studies. 

Consider the case where, for $k=1,\ldots,d$,
$
\theta^{(k)}\sim \mathcal{N}(0,\sigma_0^2)$, 
and
$y^{(k)}\mid \theta^{(k)} \sim \mathcal{N}(\theta^{(k)},\sigma_1^2)\,,
$
independently in both cases. Set $y^{(k)}=0$ and $\sigma_0^2=\sigma_1^2=1/4\pi$, 
so that $\Evid=1$ for all $d$'s.
A draw from the constrained prior is obtained 
as follows: simulate $r^2\leq -2^{1/2}\log l$ from a truncated $\chi^2(d)$ distribution and
$u_1,\ldots,u_d\sim \mathcal{N}(0,1)$, then set $\theta^{(k)}=r\,u_k/(u_1^2+\ldots+u_d^2)^{1/2}$. 
Since $\Evid=1$, we assume that the truncation point $\eps_d$ is such
that $\varphi(0)\eps_d =\tau\ll 1$, $\tau=10^{-6}$ say, 
where $\varphi(0)=2^{d/2}$ is the maximum likelihood value. 
Therefore, $\eps_d=\tau 2^{-d/2}$ and the number of iterations required to produce
a given  truncation error, that is, $j=\lceil (-\log\epsilon)N \rceil$, 
grows linearly in $d$. To assess the dependence of 
the asymptotic variance with respect to $d$, we state the following lemma, established
in Appendix 3.

\begin{lemma}\label{lem:x} 
In the current setting, if $V_d$ is the asymptotic variance of the nested 
sampling estimator with truncation point $\eps_d$, there exist constants $c_1$, $c_2$ such that
$ 
V_d/d \leq c_1 
$
~for all $d\geq 1$, and 
$ 
\liminf_{d\rightarrow + \infty} V_d/d \geq c_2.
$
\end{lemma}

This lemma is easily generalised to cases where the prior is such 
that the components are independent and identically distributed, and the likelihood
factorises as $L(\theta)=\prod_{k=1}^d L(\theta^{(k)})$. We conjecture that $V_d/d$ converges
to a finite value in all these situations and that, for more general models, the variance
grows linearly with the actual dimensionality of the problem, as measured for instance in \cite{SpiegBestCarl}.

\section{Nested importance sampling} \label{sec:nis}
We introduce an extension of nested sampling based on importance sampling. 
Let $\widetilde{\pi}(\theta)$ 
an instrumental prior with the support of $\pi$ included in  the support of $\widetilde{\pi}$,
and let $\widetilde{L}(\theta)$ an instrumental likelihood, namely
a positive measurable function. We define an importance weight function $w(\theta)$ such that
$\widetilde{\pi}(\theta)\widetilde{L}(\theta)w(\theta) = \pi(\theta)L(\theta)$.
We can approximate $\Evid$ by nested sampling for the pair
$(\widetilde{\pi}, \widetilde{L})$, that is, by simulating iteratively from $\widetilde{\pi}$
constrained to $\widetilde{L}(\theta)>l$, and by computing the generalised nested sampling estimator
\begin{equation}\label{estnis:eq}
\sum_{i=1}^{j}(x_{i-1}-x_{i})\varphi_i w(\theta_i). 
\end{equation}
The advantage of this extension is that one can choose $(\widetilde{\pi}, \widetilde{L})$ 
so that simulating from $\widetilde{\pi}$
under the constraint $\widetilde{L}(\theta)>l$ is easier than  simulating from 
$\pi$ under the constraint $L(\theta)>l$. For instance, one may choose an instrumental prior $\widetilde{\pi}$
such that Markov chain Monte Carlo steps adapted to the instrumental constrained prior are easier to implement
than with respect to the actual constrained prior.
In a similar vein, nested importance sampling facilitates contemplating several priors at once, 
as one may compute the evidence for each prior by producing the same nested sequence, based on
the same pair $(\widetilde\pi,\widetilde L)$, and by simply modifying the weight function.

Ultimately, one may choose $(\widetilde{\pi}, \widetilde{L})$  so that the constrained
simulation is performed exactly. For instance, if $\widetilde{\pi}$
is a Gaussian $\mathcal{N}_d(\hat\theta,\hat\Sigma)$ distribution
with arbitrary hyper-parameters, take
$$\widetilde{L}(\theta) = \lambda\left\{(\theta-\hat\theta)^T{\hat\Sigma}^{-1}(\theta-\hat\theta)\right\}\,,$$
where $\lambda$ is an arbitrary decreasing function. 
Then
\[
\varphi_i w(\theta_i) = \widetilde{L}(\theta_i) w(\theta_i) = 
{\pi(\theta_i)L(\theta_i)}\big/{\widetilde{\pi}(\theta_i)}\, .
\]
In this case, the $x_i$'s in \eqref{eq:RiemannSum1} are error-free: at iteration $i$, $\theta_i$ is sampled
uniformly over the ellipsoid that contains exactly $\exp(-i/N)$ prior mass as $\theta_i = q_i C v/\Vert v \Vert_2^{1/2}$,
where $C$ is the Cholesky lower triangle of $\hat\Sigma$, $v\sim N_d(0,I_d)$, and $q_i$ is the  $\exp(-i/N)$ quantile
of a $\chi^2(d)$ distribution. \cite{mukherjee2006nsa} consider a nested sampling algorithm where simulated points 
are generated within an ellipsoid, and accepted if they respect the likelihood constraint, but their algorithm is
not based on the importance sampling extension described here. 

The nested ellipsoid strategy seems useful in two scenarios. First, assume both the posterior mode and the Hessian
at the mode are available numerically and tune $\hat\theta$ and $\hat\Sigma$ accordingly. 
In this case, this strategy should outperform standard
importance sampling based on the optimal Gaussian proposal, because the nested ellipsoid strategy uses a $O(N^{-1})$ quadrature rule
on the radial axis, along which the weight function varies the most; see  \S \ref{sec:ex:nis} for an illustration. 
Second, assume only the posterior mode is available, so one may set $\hat\theta$ to the posterior mode, and set $\hat\Sigma=\tau I_d$,
where $\tau$ is an arbitrary, large value.  Section \ref{sec:ex:nis} indicates that the nested ellipsoid strategy
may still perform reasonably in such a scenario. Models such that the Hessian at the mode is tedious 
to compute include in particular Gaussian state space models with missing observations  \citep[][Chap. 12]{brockwell:davis:1996}, 
Markov modulated Poisson processes \citep{Ryden:MMP}, or, more generally, models where the expectation-maximisation algorithm
\citep[see][]{maclachlan:krishnan:1997} is the easiest way to compute the posterior mode,
although one may use Louis' (\citeyear{louis:1982}) method for computing the information matrix from the expectation-maximisation output.

\section{Alternative algorithms}\label{sec:alterego}

\subsection{Approximating $Z$ from a posterior sample}
As recalled in \S \ref{variat:sec}, the output of nested sampling can be ``recycled" so as to approximate posterior 
quantities. Conversely, one can recycle the output of an Markov chain Monte Carlo algorithm towards estimating the evidence,
with no or little additional programming effort; see for instance \cite{Gelfand:Dey:1994}, 
\cite{meng:wong:1996}, and \cite{Chen:Shao:1997}. We describe below 
the solutions used in the subsequent comparison with nested sampling, but 
we do not pretend at an exhaustive coverage of those
techniques, see \cite{chen:shao:ibrahim:2000} or \cite{han:carlin:2001}
for a deeper coverage, nor at using the most efficient approach, see \cite{meng:schilling:2002}. 

\subsection{Approximating $Z$ by a formal reversible jump}
We first recover Gelfand and Dey's (1994) solution of reverse importance sampling
by an integrated reversible jump,
because a natural approach to compute a marginal likelihood is to use a  
reversible jump Markov chain Monte Carlo algorithm \citep{Green:1995}. However, this may seem wasteful 
as it involves simulating from several models, while only one is of interest. 
But we can in theory contemplate a single model $\mathcal{M}$ and still implement reversible 
jump in the following way. Consider a formal alternative model $\mathcal{M}^\prime$, for instance a 
fixed distribution like the $\mathcal{N}(0,1)$ distribution, with prior weight $1/2$ 
and build a proposal from $\mathcal{M}$ to $\mathcal{M}^\prime$ that 
moves to $\mathcal{M}^\prime$ with probability \citep{Green:1995}
$\rho_{\mathcal{M}\rightarrow \mathcal{M}^\prime} = 
\{(1/2) g(\theta) \}\big/\{(1/2) \pi(\theta)L(\theta)\} \wedge 1
$
and from $\mathcal{M}^\prime$ to $\mathcal{M}$ with probability
$\rho_{M^\prime\rightarrow M} = 
\{(1/2) \pi(\theta)L(\theta)\}\big/\{(1/2) g(\theta) \} \wedge 1\,,
$
$g(\theta)$ being an arbitrary proposal on $\theta$. 
Were we to actually run this reversible jump Markov chain Monte Carlo algorithm, the
frequency of visits to $\mathcal{M}$ would then converge to $Z$.

However, the reversible sampler is not needed since,
if we run a standard Markov chain Monte Carlo algorithm on $\theta$ and compute the probability of
moving to $M^\prime$, the expectation of the ratio $g(\theta) 
/ \pi(\theta)L(\theta)$ is equal to the inverse of $Z$:
\[
E\left\{{g(\theta) }\big/{\pi(\theta)L(\theta)}\right\} =
\int \frac{g(\theta) }{\pi(\theta)L(\theta)} \,
\frac{\pi(\theta)L(\theta)}{Z}\,\text{d}\theta = {1}\big/{Z}\,,
\]
no matter what $g(\theta)$ is, in the spirit of both \cite{Gelfand:Dey:1994} and \cite{BartScaMi}.

Obviously, the choice of $g(\theta)$ impacts on the precision of the approximated $Z$. When
using a kernel approximation to $\pi(\theta\mid y)$ based on earlier Markov chain Monte Carlo simulations and considering
the variance of the resulting estimator, the 
constraint is opposite to the one found in importance sampling, namely that $g(\theta)$ must have 
lighter (not fatter) tails than $\pi(\theta)L(\theta)$ for the approximation 
\[
\widehat{Z_1} = 1\bigg/\left\{
\frac{1}{T}\,\sum_{t=1}^T {g(\theta^{(t)}) }\pi(\theta^{(t)})L(\theta^{(t)})\right\}
\]
to have a finite variance. This means that light tails or finite support kernels, like an
Epanechnikov kernel, are to be preferred to fatter tails kernels, like the $t$ kernel.

In the experimental comparison reported in \S \ref{Minx}, we compare $\widehat{Z_1}$ with
a standard importance sampling approximation
\[
\widehat{Z_2} = \frac{1}{T}\,\sum_{t=1}^T 
	{\pi(\theta^{(t)})L(\theta^{(t)})}\big/{g(\theta^{(t)}) }\,,
\qquad \theta^{(t)}\sim  g(\theta)\,,
\]
where $g$ can also be a
non-parametric approximation of $\pi(\theta\mid y)$, this time with heavier tails than $\pi(\theta)L(\theta)$.
\cite{fruhwirth:2004} uses the same importance function $g$ in both $\widehat{Z_1}$ 
and $\widehat{Z_2}$, and obtain results similar to ours, namely that $\widehat{Z_2}$ outperforms $\widehat{Z_1}$.

\subsection{Approximating $Z$ using a mixture representation}\label{sub:mix}

Another approach in the approximation of $Z$ 
is to design a specific mixture for simulation purposes, with density 
proportional to
$$
m(\theta)\propto \omega_1 \pi(\theta)L(\theta) + g(\theta)
$$
where $\omega_1>0$ and $g(\theta)$ is an arbitrary, fully specified density.
Simulating from this mixture has the same complexity as simulating from
the posterior, the Markov chain Monte Carlo code used to simulate from $\pi(\theta\mid y)$
can be easily extended by introducing an auxiliary
variable $\delta$ that indicates whether or not the current simulation is from $\pi(\theta\mid y)$
or from $g(\theta)$. The $t$-th iteration of this extension is as follows,
where $\mathcal{K}(\theta,\theta^\prime)$ denotes an arbitrary Markov chain Monte Carlo kernel associated with the
posterior $\pi(\theta\mid y)\propto \pi(\theta)L(\theta)$:
\begin{enumerate}
\item Take $\delta^{(t)}=1$, and $\delta^{(t)}=2$ otherwise, with probability 
\[
\omega_1 \pi(\theta^{(t-1)})L(\theta^{(t-1)})\big/ 
\left\{\omega_1 \pi(\theta^{(t-1)})L(\theta^{(t-1)}) 
+ g(\theta^{(t-1)})\right\}\,;
\]
\item If $\delta^{(t)}=1$, generate $\theta^{(t)}\sim\text{Markov chain Monte Carlo}(\theta^{(t-1)},\theta^{(t)})$, else
generate $\theta^{(t)}\sim g(\theta)$ independently from the previous value $\theta^{(t-1)}$.
\end{enumerate}
\noindent
This algorithm is a Gibbs sampler: Step 1 simulates $\delta^{(t)}$ conditional on $\theta^{(t-1)}$,
while Step 2 simulates  $\theta^{(t)}$ conditional on $\delta^{(t)}$. While the average of the 
$\delta^{(t)}$'s converges to $\omega_1 Z/\{ \omega_1 Z + 1\}$, a natural 
Rao-Blackwellisation is to take the average of the expectations of the $\delta^{(t)}$'s,
\[
\hat{\xi}=\frac{1}{T}\,\sum_{t=1}^T  \omega_1 \pi(\theta^{(t)})L(\theta^{(t)}) \bigg/  \left\{ \omega_1 
\pi(\theta^{(t)})L(\theta^{(t)}) + g(\theta^{(t)}) \right\} \,,
\]
since its variance should be smaller. A third estimate 
is then deduced from this approximation by solving $\omega_1\hat{\Evid}_3/\{\omega_1\hat{\Evid}_3+1\}=\hat{\xi}$.

The use of mixtures in importance sampling in order to improve the stability of the estimators dates back at
least to \cite{hesterberg:1998} but, as it occurs, this particular mixture estimator happens to be almost 
identical to the bridge sampling estimator of \cite{meng:wong:1996}. In fact,
\[
\hat{\Evid}_3 = \frac{1}{\omega_1} \sum_{t=1}^T  \frac{\omega_1 \pi(\theta^{(t)})L(\theta^{(t)}) }{  
\omega_1 \pi(\theta^{(t)})L(\theta^{(t)}) + g(\theta^{(t)}) } \bigg/
\sum_{t=1}^T  \frac{ g(\theta^{(t)})  }{
\omega_1 \pi(\theta^{(t)})L(\theta^{(t)}) + g(\theta^{(t)}) } 
\]
is the Monte Carlo approximation to the ratio 
$$E_m\{\alpha(\theta)\pi(\theta)L(y\mid \theta)\}/
E_m[\alpha(\theta)g(\theta)]$$
when using the optimal function 
$
\alpha(\theta) = 1 \big/ \{\omega_1 \pi(\theta)L(\theta) + g(\theta)\}\,.
$
The only difference with \cite{meng:wong:1996} is that, since $\theta^{(t)}$'s 
are simulated from the mixture, they can be recycled for both sums.

%

%

\section{Numerical experiments}  \label{sec:simus}

\subsection{A decentred Gaussian example}\label{Cantbury} \label{decentred:sec}

We modify the Gaussian toy example presented in \S \ref{sec:toy}: 
$\theta=(\theta^{(1)},\ldots,\theta^{(d)})$, where the $\theta^{(k)}$'s are
independent and identically distributed from $\mathcal{N}(0,1)$, 
and $y_k\mid \theta^{(k)} \sim \mathcal{N}(\theta^{(k)},1)$ independently, but setting
all the $y_k$'s to $3$. To simulate from the prior truncated to 
$L({\theta})>L({\theta}_0)$, we perform $M$ Gibbs iterations 
with respect to this truncated distribution, with $M=1$, 3 or 5: 
the full conditional distribution of $\theta^{(k)}$, conditional on $\theta^{(j)}$, $j\neq k$, is a
$\mathcal{N}(0,1)$ distribution that is truncated to the interval 
$[y^{(k)}-\delta,y^{(k)}+\delta]$ with 
\[
\delta^2=\sum_j (y_j-\theta_0^{(j)})^2-\sum_{j\neq k} (y_j-\theta^{(j)})^2.
\]

\begin{figure}
\centerline{\includegraphics[keepaspectratio,width=16cm]{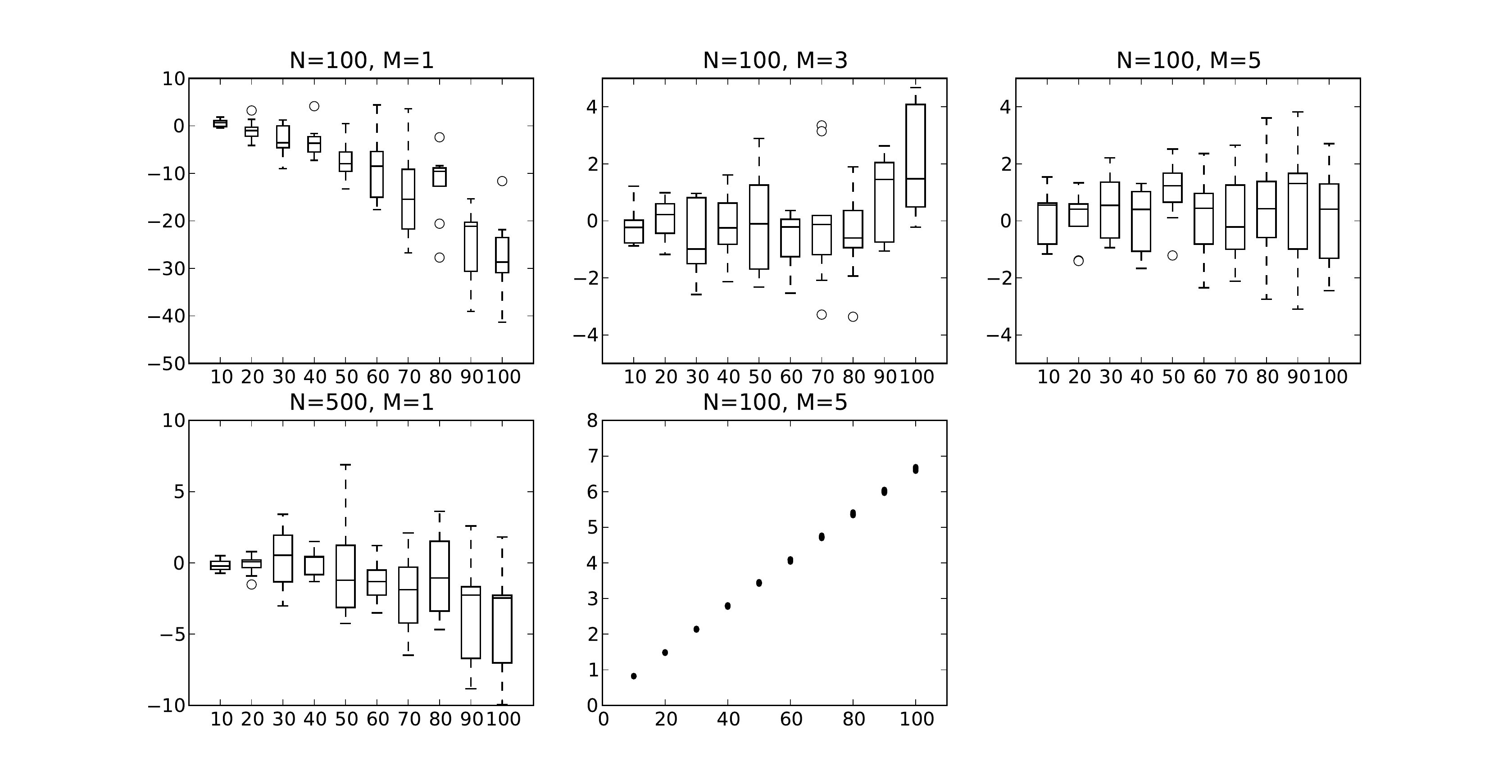}}
\caption{\label{fig:decGauss} Decentred Gaussian example: Box-plots of the log-relative error 
$\log \widehat\Evid-\log \Evid$ versus dimension $d$ for four values of $(N,M)$, 
and (lower right) total number of iterations $(\times 10^4)$ versus dimension for $(N,M)=(100,5)$
}
\end{figure}

The nested sampling algorithm is run $20$ times for 
$d=10$, $20$, $\ldots$, $100$, and several combinations of $(N,M)$: $(100,1)$, $(100,3)$, $(100,5)$, and $(500,1)$. 
The algorithm is stopped when a new contribution 
$(x_{i-1}-x_{i})\varphi_i$ to \eqref{eq:RiemannSum1}
becomes smaller than $10^{-8}$ times the current estimate. 
Focussing first on $N=100$, Fig.~\ref{fig:decGauss} exposes the impact of the mixing properties of the Markov chain Monte Carlo step: 
for $M=1$, the bias sharply increases with respect to the dimension, while, for $M=3$, it remains small 
for most dimensions.  Results for $M=3$ and $M=5$ are quite similar, except perhaps for $d=100$. 
Using $M=3$ Gibbs steps seems to be sufficient to produce a good approximation
of an ideal nested sampling algorithm, where points would be independently simulated.
Interestingly, if $N$  increases to $500$, while keeping $M=1$, then 
larger errors occur for the same computational effort. Thus, a good strategy in this case is
to increase first $M$ until the distribution of the error stabilises, then to increase $N$
to reduce the Monte Carlo error. 
As expected, the number of iterations linearly increases with the dimension.

While artificial, this example shows that nested sampling may perform
quite well even in large dimension problems, provided $M$ is large enough. 

\subsection{A mixture example}\label{Minx}
\begin{figure}
 \centerline{\includegraphics[height=5truecm,width=10cm]{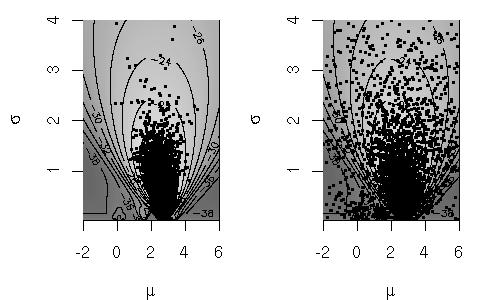}}
 \caption{\label{fig:MCMC1} Mixture example: {\em (left)} Markov chain Monte Carlo sample 
 plotted on the log-likelihood surface in the $(\mu,\sigma)$ space for $n=10$ observations from
 \eqref{eq:mix} {\em (right)} nested sampling sequence based on
 $N=10^3$ starting points for the same dataset}
 \end{figure}

As in \cite{fruhwirth:2004}, we consider the example of the posterior
distribution on $(\mu,\sigma)$ associated with the normal mixture
\begin{equation}\label{eq:mix}
y_1,\ldots,y_n\sim p\mathcal{N}(0,1)+(1-p)\mathcal{N}(\mu,\sigma)\,,
\end{equation}
when $p$ is known, for two compelling reasons. First, 
when $\sigma$ converges to $0$ and $\mu$ is equal to any of the $x_i$'s $(1\le i\le n)$, the likelihood
diverges, see Fig.~\ref{fig:MCMC1}. This is a priori challenging for exploratory schemes such as nested sampling.
Second, efficient Markov chain Monte Carlo strategies have been developed for mixture models
\citep{Diebolt:Robert:1994, Richardson:Green:1997,Celeux:Hurn:Robert:2000}, but
Bayes factors are difficult to approximate in this setting.

We simulate $n$ observations from a $\mathcal{N}(2,(3/2)^2)$
distribution, and then compute the estimates of $\Evid$ introduced above for the model \eqref{eq:mix}.
The prior distribution is uniform  on $(-2,6)\times(0\cdot001,16)$ for $(\mu,\log\sigma^2)$. 
The prior is arbitrary, but it allows for an easy implementation of
nested sampling since the constrained simulation can be implemented via a random walk move.

The two-dimensional nature of the parameter space
allows for a numerical integration of $L(\theta)$, based on a Riemann approximation and a grid
of $800\times500$ points in the $(-2,6)\times(0\cdot001,16)$ square. This approach leads to a stable evaluation
of $\Evid$ that can be taken as the reference against which we can test the various methods, since additional
evaluations based on a crude Monte Carlo integration 
using $10^6$ terms and on \citeauthor{chib:1995}'s (1995) produced essentially the same numerical values.
The Markov chain Monte Carlo algorithm implemented here is the standard completion of \cite{Diebolt:Robert:1994},
but it does not suffer from the usual label switching deficiency \citep{jasra:holmes:stephens:2005} 
because \eqref{eq:mix} is identifiable.
As shown by the Markov chain Monte Carlo sample of size $N=10^4$ displayed on the left hand side of Fig.~\ref{fig:MCMC1},
the exploration of the modal region by the Markov chain Monte Carlo chain is satisfactory.
This Markov chain Monte Carlo sample is used to compute the non-parametric approximations $g$ 
that appear in the three alternatives of \S \ref{sec:alterego}. For the reverse 
importance sampling estimate $\Evid_1$, $g$ is a product of two Gaussian kernels
with a bandwidth equal to half the default bandwidth of the R function density(), while, for 
both $\Evid_2$ and $\Evid_3$, $g$ is a product of two $t$ kernels with a
bandwidth equal to twice the default Gaussian bandwidth.
 
We ran the nested sampling algorithm, with $N=10^3$, reproducing the implementation of \cite{Skilling:2007a}, namely
using $10$ steps of a random walk in $(\mu,\log\sigma)$ constrained by the likelihood boundary.
based on the contribution of the
current value of $(\mu,\sigma)$ to the approximation of $\Evid$. The overall number of points produced by nested sampling
at stopping time is on average close to $10^4$, which justifies using the same number of points for the Markov chain Monte Carlo algorithm.
As shown on the right hand side of Fig.~\ref{fig:MCMC1}, the nested sampling sequence visits the minor modes of the likelihood surface but
it ends up in the same central mode as the Markov chain Monte Carlo sequence. 
All points visited by nested sampling are represented without reweighting, which explains for 
a larger density of points outside the central modal region.

\begin{figure}
 \centerline{\includegraphics[height=4.9truecm]{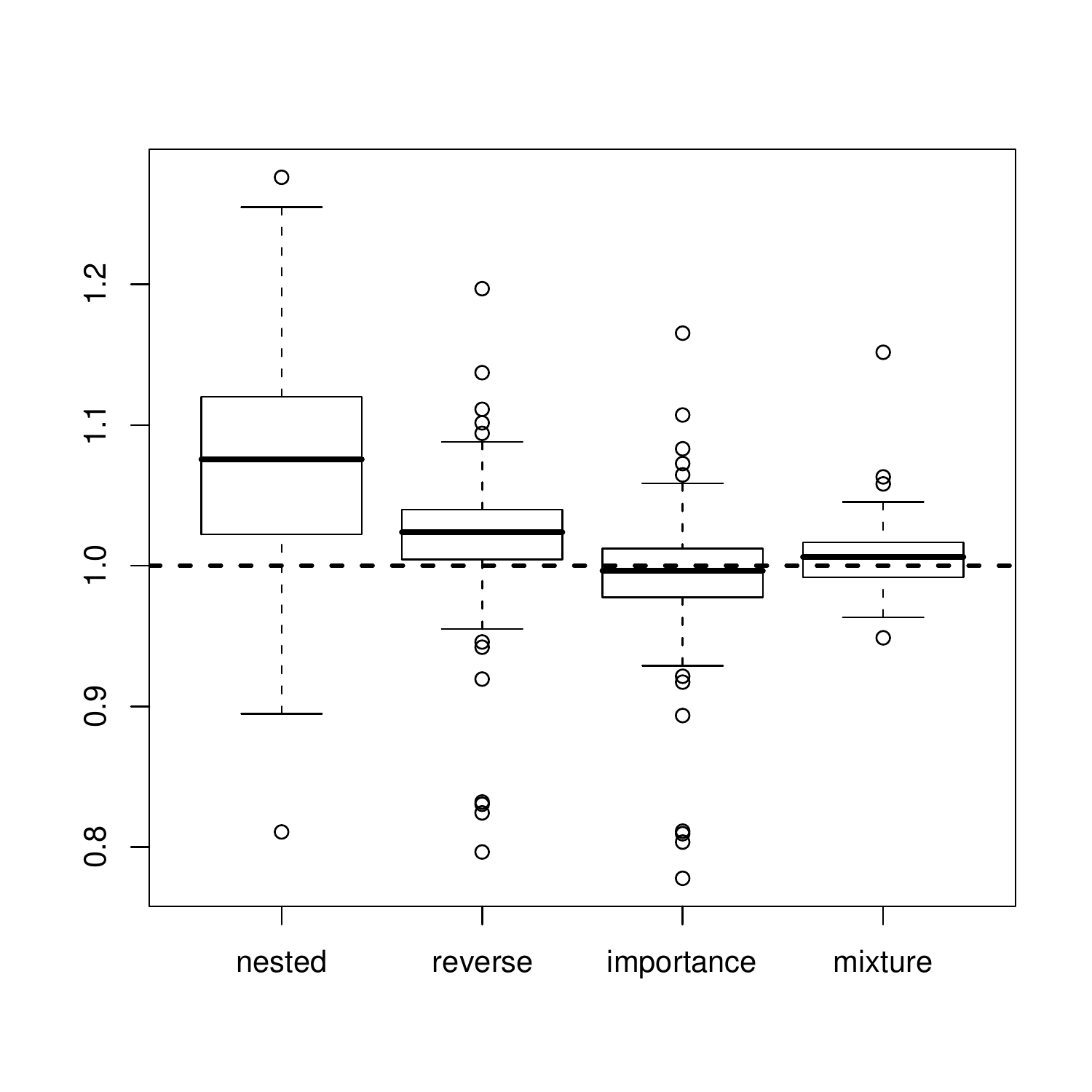}
 \includegraphics[height=4.9truecm]{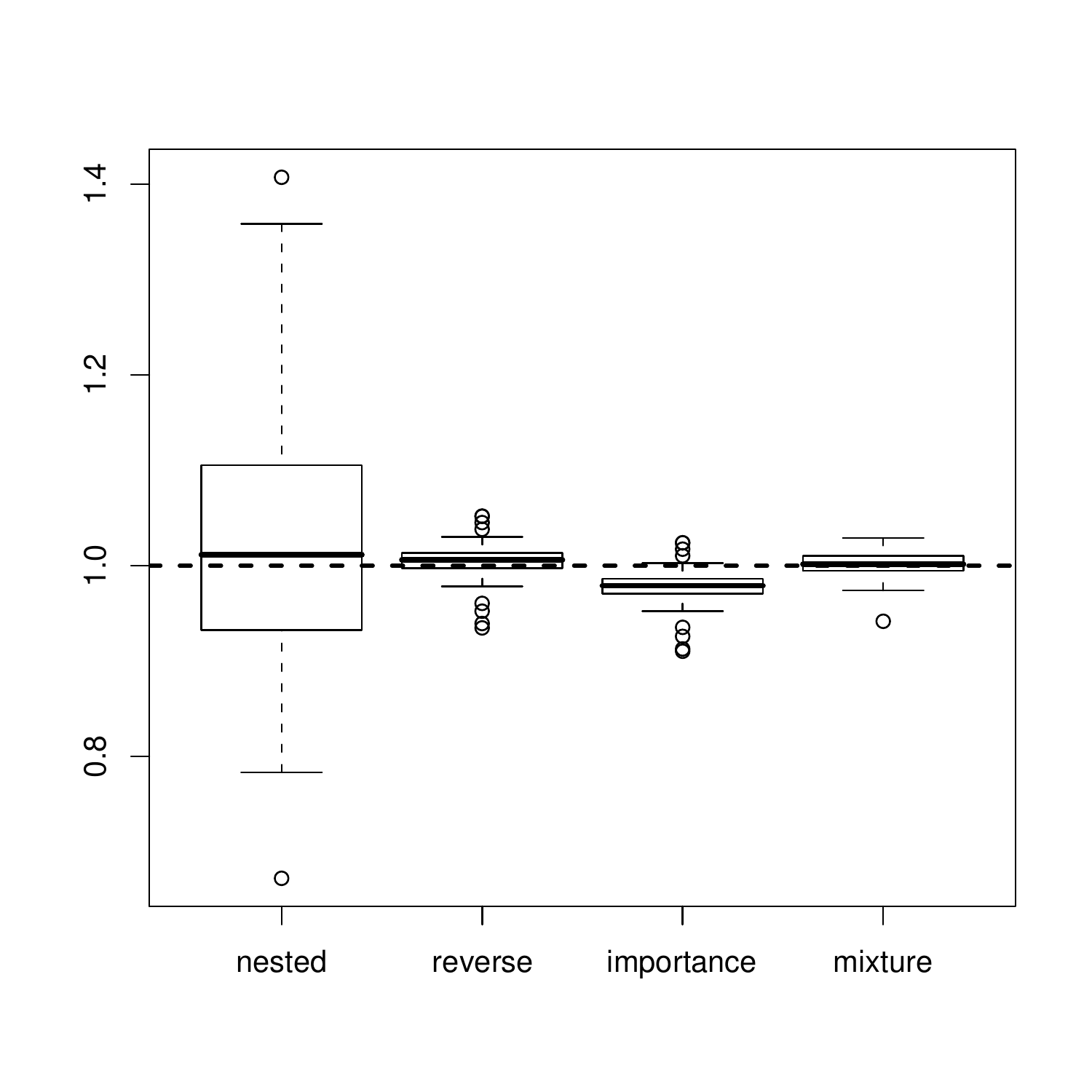}\includegraphics[height=4.9truecm]{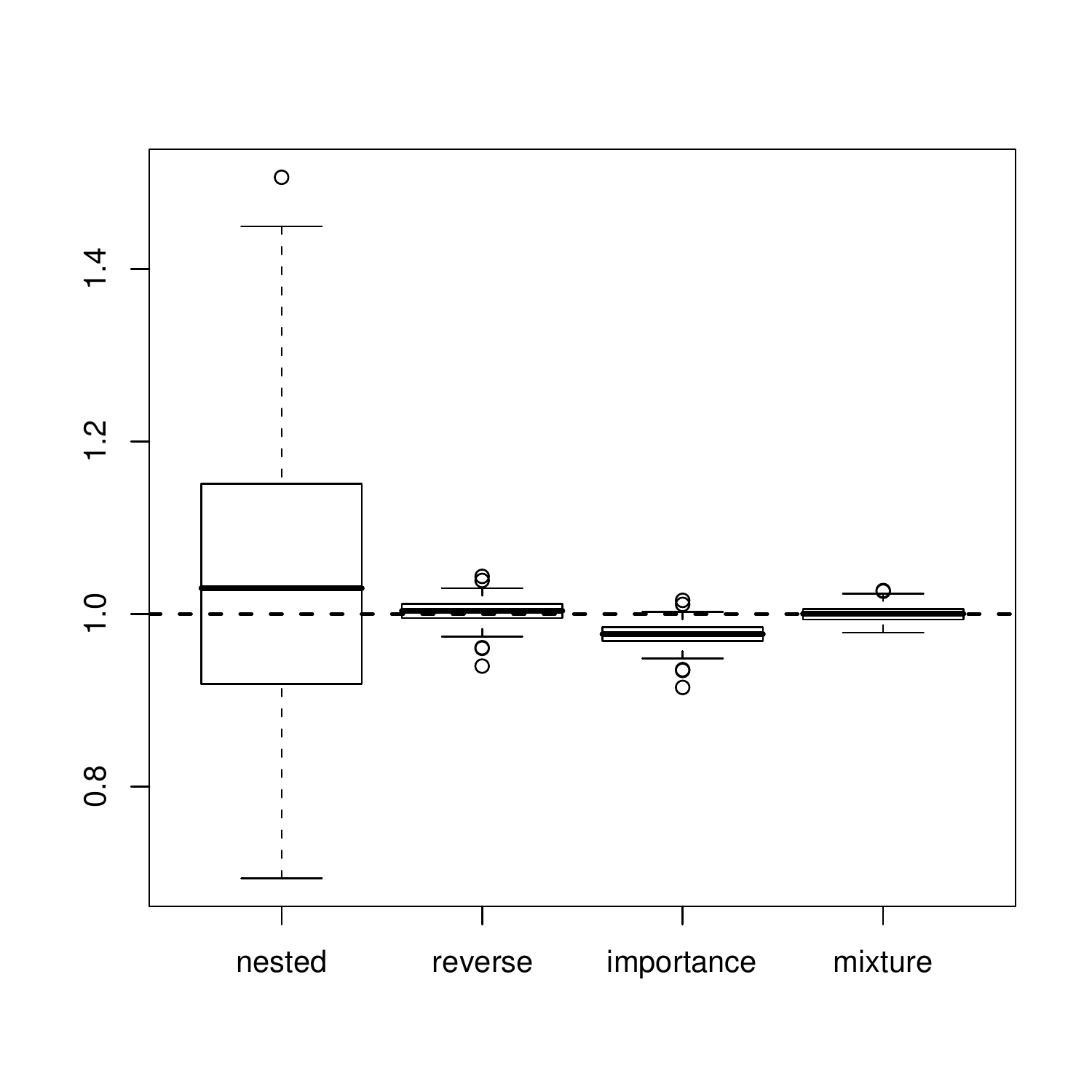}}
 \caption{\label{fig:box10} Mixture model: comparison of the variations of nested sampling,
 reverse importance sampling, importance sampling and mixture sampling, relative to
 a numerical approximation of $\Evid$ (dotted line), based on $150$ samples of size $n=10,50,100$}
 \end{figure}

The analysis of this Monte Carlo experiment in Fig.~\ref{fig:box10}
first shows that nested sampling gives approximately the same numerical value 
when compared with the three other approaches, exhibiting a slight upward bias,
but that its variability is higher. The most reliable approach, besides the numerical and raw
Monte Carlo evaluations which cannot be used in general settings, is the importance sampling solution, 
followed very closely by the mixture approach of \S \ref{sub:mix}. The reverse importance sampling 
naturally shows a slight upward bias for the smaller values of $n$ 
and a variability that is very close to both other alternatives, especially for larger values of $n$.

\subsection{A probit example for nested importance sampling} \label{sec:ex:nis}

To implement the nested importance sampling algorithm based on nested ellipsoids, 
we consider the arsenic dataset and a probit model studied in Chapter 5
of \cite{Gelman:Hill:2006}. The observations are independent Bernoulli variables $y_i$
such that $\/ext{pr}(y_i=1\mid x_i)=\Phi(x_i^T\theta)$, where $x_i$ is a vector of $d$ covariates,
$\theta$ is a vector parameter of size $d$, and $\Phi$
denotes the standard normal distribution function. In this particular example,
$d=7$; more details on the data and the covariates are available on the
book's web-page
({\small{\verb+http://www.stat.columbia.edu/~gelman/arm/examples/arsenic+}}).

The probit model we use is model 9a in the R program available 
at this address: the dependent variable indicates whether or not the surveyed individual
changed the well she drinks from over the past three years,
and the seven covariates are an intercept, distance to the nearest safe well (in 100 metres unit), 
education level, log of arsenic level, and cross-effects for these three variables. 
We assign $\mathcal{N}_d(0,10^2 I_d)$ as our prior on $\theta$, and denote $\theta_{m}$ the posterior mode, 
and $\Sigma_{m}$ the inverse of minus twice the Hessian at the mode; both quantities are obtained numerically beforehand. 

\begin{figure}[h]
\centerline{
\includegraphics*[clip=true,scale=0.6]{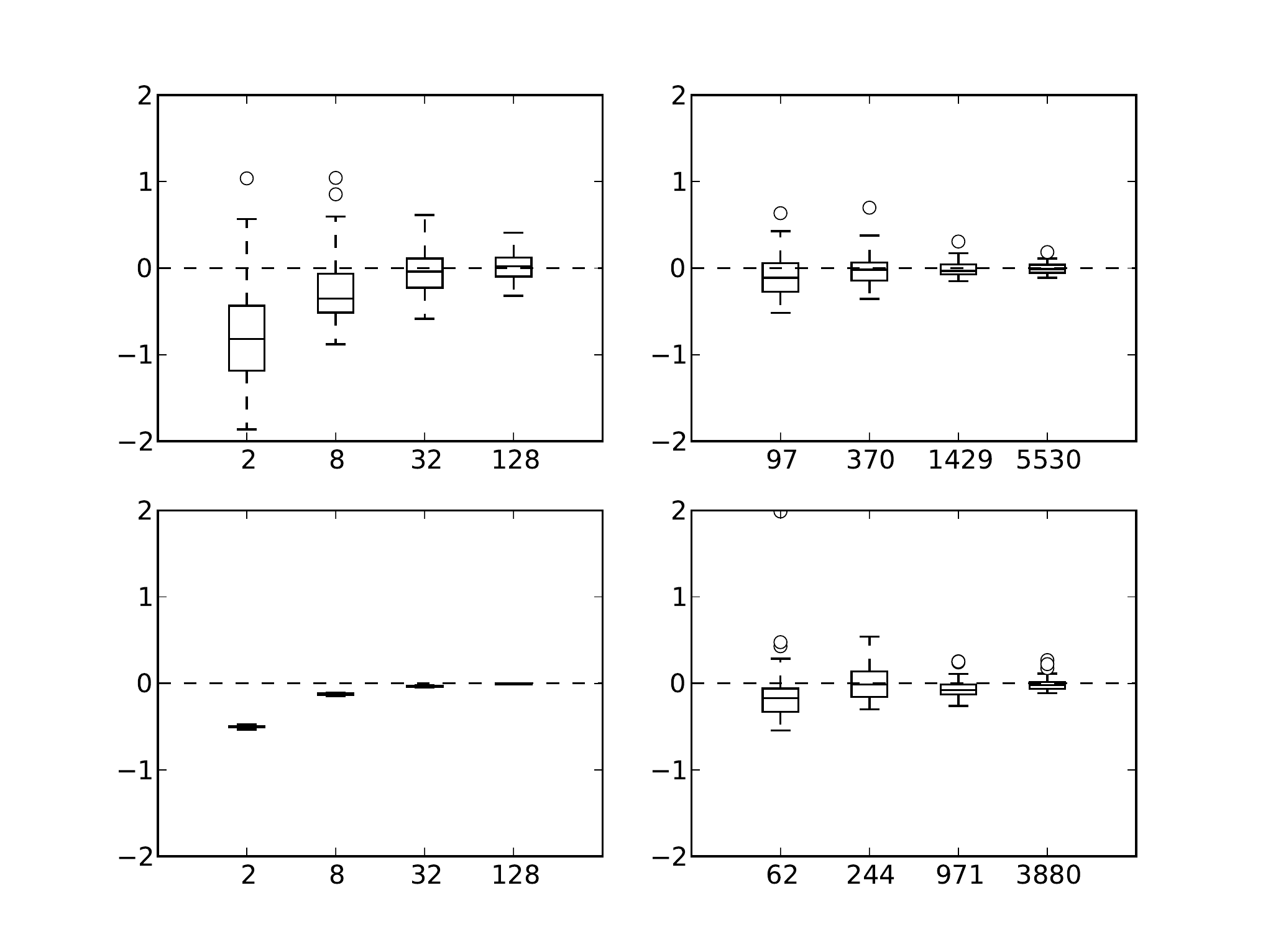}
}
\caption{\label{fig:probit} Probit example: Box-plots of (left column) log-errors of nested importance
sampling estimates, for $N=2,$ $8,$ $32$, $128$, compared with the log-error of importance sampling estimates 
(right column) based on the optimal Gaussian proposal, and the same number of likelihood evaluations. 
Those are reported on the x axis of the right column plots. The bottom row corresponds to 
the first strategy, based on both mode and Hessian, while the top row corresponds to the second strategy, based on mode only.}
\end{figure}

We run the nested ellipsoid algorithm 50 times, for $N=2$, 8, 32, 128, and for two sets
of hyper-parameters corresponding to both scenarios described in \S\ref{sec:nis}. In the first scenario,
$(\hat\theta,\hat\Sigma)=(\theta_m,2\Sigma_m)$. 
The bottom row of Fig.~\ref{fig:probit} compares log-errors produced by our method (left), with those of 
importance sampling based on the optimal Gaussian proposal, with mean $\theta_m$, variance $\Sigma_m$, 
and the same number of likelihood evaluations, as reported on the x-axis of the right plot. 
In the second scenario, $(\hat\theta,\hat\Sigma)=(\theta_m,100\, I_d)$. 
The top row of Fig.~\ref{fig:probit} compares log-errors produced by our method (left) with those of 
importance sampling, based again on the optimal proposal, and the same  number of likelihood evaluations. 
The variance of importance sampling estimates based on a Gaussian proposal with hyper-parameters $\hat\theta$ and 
 $\hat\Sigma=100 I_d$ is higher by several order of magnitudes, and is not reported in the plots. 

As expected, the first strategy outperforms standard importance sampling, when both methods are supplied
with the same information (mode, Hessian), and the second strategy still does reasonably well 
compared to importance sampling based on the optimal Gaussian proposal, although only provided with the mode.
Results are sufficiently precise that one can afford to compute the evidence for the $2^7$ possible models:  
the most likely model, with posterior probability $0.81$, includes the 
intercept, the three variables mentioned above, distance, arsenic, education, and one cross-effect 
between distance and education level, and the second most likely model, with posterior probability $0.18$,  
is the same model but without the cross-effect. 

\section{Discussion}
Nested sampling is thus a valid addition to the Monte Carlo toolbox,
with convergence rate $\text{O}(N^{-1/2})$, and computational cost $O(d^3)$, where $d$ is the dimension of the problem. 
which enjoys good performances in some applications, for example when the posterior is approximately
Gaussian, but which may require more iterations to achieve the same precision in certain situations. 
Therefore, further work on the formal and practical assessments of nested sampling convergence would be welcome. For one
thing, the convergence properties of Markov chain Monte Carlo-based nested sampling are unknown and technically challenging.
Methodologically, efforts are required to design efficient Markov chain Monte Carlo moves with respect to 
the constrained prior. In that and other respects, nested importance sampling may constitute a useful extension. Ultimately,
our comparison between nested sampling and alternatives should be extended to more diverse examples, in order to get
a clearer idea of when nested sampling should be the method of choice and when it should not. 
For instance, \cite{Murray:nested:Potts} reports that nested sampling strongly outperforms annealed 
importance sampling \citep{neal2001ais} for Potts models. All the programs implemented for this paper are available from the authors.

\section*{Acknowledgement}
The authors are grateful to 
R. Denny, A. Doucet, T. Loredo,
O. Papaspiliopoulos,  G. Roberts, 
J. Skilling, the Editor, the Associate Editor and the referees for helpful comments. 
The second author is also a member of the Center for Research in Economy and Statistics (CREST),
whose support he gratefully acknowledges.

\bibliography{cur.biblio}
\bibliographystyle{biometrika}

\appendix
\section*{Appendix 1}\subsection*{Proof of Lemma \ref{post:Lemma}}

It is sufficient to prove
this result for functions $\widetilde{f}$ that are real-valued, positive and increasing.
First, the extension to vector-valued functions is trivial, so 
$\widetilde{f}$ is assumed to be real-valued from now on. Second, the class of functions that satisfy
property (\ref{eq:postequ})  is clearly stable through addition.  Since $\widetilde{f}$ 
is  absolutely continuous, there exist functions $f^+$ and $f^-$,
such that $f^+$ is increasing, $f^-$ is decreasing, and $\widetilde{f}=f^++f^-$,
so we can restrict our attention to increasing functions. Third,
absolute continuity implies bounded variation, so it always possible
to add an arbitrary constant to  $\widetilde{f}$ to transform it into a positive function. 

Let $\psi:l\rightarrow l\widetilde{f}(l)$, which is a positive, increasing function
and denote its
inverse by $\psi^{-1}$. One has:
\[
   \E^{\pi} [ \psi\{L(\theta)\} ]
  =  \int_0^{+\infty} \mathrm{pr}[ \psi\{L(\theta)\}>l ]\,\text{d}l 
 =  \int_0^{+\infty} \varphi^{-1}\{\psi^{-1} (l) \}\,\text{d}l 
  =   \int_0^1 \psi\{\varphi(x)\}\,\text{d}x\,,
\]
which concludes the proof. 

\section*{Appendix 2}\subsection*{Proof of Theorem 1}

Let $t_{i}=x_{i+1}^{\star}/x_{i}^{\star}$, for $i=0,1,\ldots$ As mentioned by \cite{Skilling:2007a},
the $t_{i}$'s are independent beta$(N,1)$ variates. 
Thus, $u_{i}=t_{i}^{N}$ defines a sequence of independent uniform $[0,1]$ variates. A Taylor expansion of
$\eta_N$ gives:
\begin{eqnarray*}
\eta_N & = & \sum_{i=1}^{\left\lceil cN\right\rceil }(x_{i-1}-x_{i})
	\left\{\varphi(x_{i}^{\star})-\varphi(x_{i})\right\}\\
 & = & \sum_{i=1}^{\left\lceil cN\right\rceil }(x_{i-1}-x_{i})
	\left\{\psi'(-\log x_{i})\left(\log x_{i}-\log x_{i}^{\star}\right)
	+O\left(\log x_{i}-\log x_{i}^{\star}\right)^{2}\right\}
\end{eqnarray*}
where $c=-\log\eps$, and $\psi(y)=\varphi(e^{-y})$. Furthermore,
\[
S_{i}=N\left(\log x_{i}-\log x_{i}^{\star}\right)=\sum_{k=0}^{i-1}(-1-\log u_{k})
\]
is a sum of independent, standard variables, as $E(\log u_{i})=-1$
and $\text{var}(\log u_{i})=1$. Thus, $\left(\log x_{i}-\log x_{i}^{\star}\right)=\text{O}_{P}(N^{-1/2})$,
where the implicit constant in $\text{O}_{P}(N^{-1/2})$ does not depend on $i$,
and 
\begin{eqnarray*}
N^{1/2}\eta_N & = & N^{-1/2}\sum_{i=1}^{\lceil cN \rceil}(e^{-(i-1)/N}-e^{-i/N})
	S_{i}\left\{\psi'(\frac{i}{N})+O_P(N^{-1/2})\right\}\\
 & = & c^{1/2}\sum_{i=1}^{\lceil cN \rceil}\int_{(i-1)/N}^{i/N}e^{-t}\psi'(t)B_{N}
	(\frac{t}{c})\  \text{d}t\left\{1+O_P(N^{-1/2})\right\}\,,
\end{eqnarray*}
since $\psi'(t)=\psi'(i/N)+\text{O}(N^{-1})$ for $t\in[(i-1)/N,i/N]$, where, again, 
the implicit constant in $\text{O}(N^{-1})$ can be the same for all $i$, as $\psi''$ is bounded, 
and provided $B_{N}(t)$ is defined as
$
B_{N}(t)=(cN)^{-1/2}S_{\left\lceil cNt\right\rceil }
$
for $t\in[0,1]$. According to Donsker's theorem \citep[][p.275]{Kallenberg:2002}, 
$B_{N}$ converges to a Brownian motion $B$ on $[0,1]$,
in the sense that $f(B_{N})$ converges in distribution to $f(B)$ for any measurable and
a.s. continuous function $f$. Thus
\[
N^{1/2}\eta_N  = c^{1/2}\int_{0}^{\left\lceil cN\right\rceil /N}e^{-t}\psi'(t)B_{N}
	(\frac{t}{c})\  \text{d}t+O_P(N^{-1/2})
\]
converges in distribution to 
$$c^{1/2}\int_{0}^{c}e^{-t}\psi'(t)B(\frac{t}{c})\, \text{d}t\,,$$
which has the same distribution as the following zero-mean Gaussian variate: 
\[
\int_{0}^{c}e^{-t}\psi'(t)B(t)\, \text{d}t=\int_{\eps}^{1}s\varphi'(s)B(-\log s)\, \text{d}s.
\]

\section*{Appendix 3}\subsection*{Proof of Lemma \ref{lem:x}}

For the sake of clarity, we make dependencies on $d$ explicit in this section, 
including $\varphi_d$ for $\varphi$, $\eps_d$ for $\eps$, and so on. We will use
repeatedly the facts that $\varphi$ is nonincreasing and that $\varphi'$ is 
nonnegative. One has: 
$$
-\int_{s,t\in[\varepsilon_{d},1]}s\varphi_d'(s)t\varphi_d'(t)\log(s\vee t)\,\text{d}t  \leq  -\log\varepsilon_{d}\left\{\int_{\varepsilon_{d}}^{1}s\varphi_d'(s)\,\text{d}s\right\}^{2}  
\leq d\log(2^{1/2}/\tau)
$$
for $d\geq1$, since $-\int_{\varepsilon_{d}}^{1}s\varphi_d'(s)\,\text{d}s\leq-\int_{0}^{1}s\varphi_d'(s)\,\text{d}s=
1$. This gives the first result.

Let $s_d = \varphi_d^{-1} ( \alpha^d)$, for $0<\alpha<1$; $s_d$  is the probability that 
\[ 
(4\pi/d){\sum_{i=1}^d \theta_i^2} -1 \leq - 2 \log (\alpha)+\log(2) -1 
\]
assuming that the $\theta_i$'s are independent $\mathcal{N}(0,1/4\pi)$ variates. The
left-hand side is an empirical average of independent and identically distributed~zero-mean variables. We take $\alpha$
so that the right-hand side is negative, which implies $\alpha > 2^{1/2} \exp(-1/2)$. 
Using large deviations \citep[Chapter 27]{Kallenberg:2002}, 
one has $-\log(s_d)/d\rightarrow \gamma>0$ as $d\rightarrow + \infty,$ and
\begin{eqnarray*}
\frac 1 d V_d &=& 
-\frac 1 d \int_{s,t\in[\eps_d,1] } s\varphi_d'(s)t\varphi_d'(t) \log (s \vee t) \, \text{d}s \text{d}t \\
& \geq & \left\{ \frac{-\log s_d}{d}\right) \left( \int_{\eps_d}^{s_d} s\varphi_d'(s)\, \text{d}s\right\}^2\\
& \geq &  \left( \frac{-\log s_d}{d}\right) \left\{
\int_{\eps_d}^{s_d} \varphi_d(s)\, \text{d}s 
+\eps_d \varphi_d(\eps_d) - s_d\varphi_d(s_d)
\right\}^2 \\
& \geq & \left( \frac{-\log s_d}{d}\right) \left\{
1-\int_0^{\eps_d} \varphi_d(s)\, \text{d}s 
-\int_{s_d}^1 \varphi_d(s)\, \text{d}s  
 +\eps_d \varphi_d(\eps_d)-s_d\varphi_d(s_d)
\right\}^2 .
\end{eqnarray*}
As $d\rightarrow +\infty$, $-\log(s_d)/d \rightarrow \gamma$, 
$s_d\rightarrow 0$, $\varphi_d(s_d)=\alpha^d\rightarrow 0$,  
$\int_{s_d}^1 \varphi_d(s)\,\text{d}s\leq \varphi_d(s_d) (1-s_d)\rightarrow 0$, and 
\[ 0 \leq  \int_{0}^{\eps_d} \varphi_d(s)\, \text{d}s 
- \eps_d \varphi_d(\eps_d) \leq  \eps_d
\{\varphi_d(0) -\varphi_d(\eps_d)\} \leq \tau <1, \]
by the definition of $\eps_d$, and the squared factor is in the 
limit greater than or equal to $(1-\tau)^2$. 

\end{document}